\documentclass[aps,prl,twocolumn,nofootinbib,preprintnumbers]{revtex4}
\usepackage[dvips]{graphicx}
\usepackage{bm,latexsym,amsmath,amssymb,amsfonts,mathrsfs}
\usepackage{color}
\input{colordvi.tex}
\newcommand*{\D}{{\rm d}}
\newcommand*{\mpl}{M_{\rm Pl}}
\begin{document}

\title{New Cosmological Solutions in Massive Gravity}

\author{Tsutomu Kobayashi}
\email[Email: ]{tsutomu"at"rikkyo.ac.jp}
\affiliation{Department of Physics, Rikkyo University, Toshima, Tokyo 175-8501, Japan
}

\author{Masaru Siino}
\email[Email: ]{msiino"at"th.phys.titech.ac.jp}
\affiliation{Department of Physics, Tokyo Institute of Technology, Tokyo 152-8551, Japan}

\author{Masahide Yamaguchi}
\email[Email: ]{gucci"at"phys.titech.ac.jp}
\affiliation{Department of Physics, Tokyo Institute of Technology, Tokyo 152-8551, Japan}

\author{Daisuke Yoshida}
\email[Email: ]{yoshida"at"th.phys.titech.ac.jp}
\affiliation{Department of Physics, Tokyo Institute of Technology, Tokyo 152-8551, Japan}

\begin{abstract}
We find new, simple cosmological solutions with flat, open, and
closed spatial geometries, contrary
to the previous wisdom that only the open model is allowed.
The metric and the St\"{u}ckelberg fields are given explicitly,
showing nontrivial configurations of the St\"{u}ckelberg
in the usual Friedmann-Lema\^{i}tre-Robertson-Walker coordinates.
The solutions exhibit self-acceleration, while being free from
ghost instabilities.
Our solutions can accommodate inhomogeneous dust collapse
represented by the Lema\^{i}tre-Tolman-Bondi metric as well.
Thus, our results can be used not only
to describe homogeneous and isotropic cosmology
but also to study gravitational collapse in massive gravity.
\end{abstract}

\preprint{RUP-12-3}
\maketitle

It is very intriguing to explore whether or not the graviton {\em can}
have a mass.  The first attempt to add a mass term to the gravity action
was made by Fierz and Pauli~\cite{FP}, who considered the quadratic
action for the graviton $h_{\mu\nu}$ in flat space with the mass term
\begin{eqnarray}
m^2\left(h_{\mu\nu}h^{\mu\nu}-h^2\right).
\end{eqnarray}
The linear theory with the Fierz-Pauli mass term is ghost-free.
However, the theory does not reproduce general relativity
in the massless limit $m\to 0$.
The extra three degrees of freedom in a massive spin 2 survive even in this limit,
and therefore the prediction for light bending is away from that of general relativity,
which clearly contradicts solar-system tests.
This is called the vDVZ discontinuity~\cite{vDVZ}.

As pointed out by Vainshtein~\cite{Vainshtein},
the discontinuity can in fact be cured by going beyond the linear theory.
Massive gravity has a new length scale called the Vainshtein radius,
below which the nonlinearities of the theory come in and
the effect of the extra degrees of freedom is screened safely.
The Vainshtein radius
becomes larger as $m$ gets smaller,
and thereby a smooth massless limit is attained.

However, the very nonlinearities turned out to cause another trouble.
Boulware and Deser argued that
there appears a sixth scalar degree of freedom at nonlinear order,
which has a wrong sign kinetic term, {\em i.e.,}
the sixth mode is a ghost~\cite{BDGhost}.
The ghost issue was emphasized
in the effective field theory approach in Ref.~\cite{eft}.
The presence of the Boulware-Deser (BD) ghost has hindered us
from constructing a consistent theory of massive gravity.

Recently, a theoretical breakthrough in this field has been made.
Adding higher-order self-interaction terms and tuning appropriately
their coefficients, de Rham and collaborators successfully eliminated
the dangerous scalar mode from the theory in the decoupling limit~\cite{dR1, dR2}.
Then, Hassan and Rosen established a complete proof that the theory
does not suffer from the BD ghost instability
to all orders in perturbations and away from the decoupling limit~\cite{HR}.
Thus, there certainly exists a nonlinear theory of massive gravity
that is free of the BD ghost.

In addition to the theoretical interests described above, the
mystery of the accelerated expansion of the Universe~\cite{sne}
motivates massive gravity theories as a possible alternative to dark
energy. Since the attractive force mediated by a massive graviton is
Yukawa-suppressed by a factor $e^{-mr}$, massive gravity theories with
$m\sim H_0$ (the present Hubble rate) could help if one were to avoid
dark energy.  Indeed, the DGP model~\cite{DGP}, a concrete realization
of massive gravity in the context of extra dimensions, admits a
self-accelerating solution without the need of dark energy~\cite{DGP2}.


Then, one may wonder whether or not the massive gravity theory
developed by de Rham and collaborators~\cite{dR1, dR2} can admit flat
Friedmann-Lema\^{i}tre-Robertson-Walker (FLRW) cosmology.
Starting from the usual FLRW metric ansatz, it has been
shown that a spatially flat solution is prohibited in the above massive
gravity theory~\cite{Cos}. Though the same argument applies to the
closed model as well, this interesting fact does not hold true for the
open model, and indeed the open FLRW solution has been obtained in
Ref.~\cite{Open1}. The conclusion, however, depends upon the form of the
St\"{u}ckelberg fields one chooses. In other words, one can start from a
nonstandard form of the cosmological metric in the unitary gauge, and
then move to the usual FLRW coordinates with a nontrivial form of the
St\"{u}ckelberg fields, which would lead one to different conclusions.

In this {\em Letter}, we show that flat, closed, and open
cosmological solutions can indeed be realized even in massive gravity,
starting from a general Painlev\'{e}-Gullstrand (PG)
metric~\cite{gPG,Kanai} in the unitary gauge. The key trick used
is that the FLRW metric can be recast in a (less familiar) PG
form~\cite{PG}. Our new solutions also include a
spherical, inhomogeneous dust collapse model described by the
Lema\^{i}tre-Tolman-Bondi
(LTB) solution represented in the PG-type coordinates.
Thus, our solutions can accommodate not only the flat
cosmological model\footnote{Though
inflation does not necessarily predict an exactly flat
universe, such a solution is useful for describing our Universe and
investigating the cosmological perturbations.} but also gravitational
collapse solutions.
Many other interesting
solutions expressed in the PG coordinates
in general relativity are also solutions
to massive gravity.


The nonlinear massive gravity theory we consider is described by the
action~\cite{dR1, dR2}
\begin{eqnarray}
S=\frac{\mpl^2}{2}\int\D^4x\sqrt{-g}\left(R+m^2{\cal U}\right)+S_{\rm m},
\end{eqnarray}
where $R$ is the Einstein-Hilbert term and
${\cal U}$ is the potential for the graviton,
\begin{eqnarray}
{\cal U}:={\cal U}_2+\alpha_3{\cal U}_3+\alpha_4{\cal U}_4,
\end{eqnarray}
with two free parameters $\alpha_3$ and $\alpha_4$,
in addition to the graviton mass $m$.
Each term is defined as
\begin{eqnarray}
{\cal U}_2&:=&[{\cal K}]^2-[{\cal K}^2],
\\
{\cal U}_3&:=&[{\cal K}]^3-3[{\cal K}][{\cal K}^2]+2[{\cal K}^3],
\\
{\cal U}_4&:=&[{\cal K}]^4-6[{\cal K}^2][{\cal K}]^2+
8[{\cal K}^3][{\cal K}]+3[{\cal K}^2]^2-6[{\cal K}^4],\;\;\;
\end{eqnarray}
where
\begin{eqnarray}
{\cal K}_{\mu}^{\;\nu}:=\delta_{\mu}^{\;\nu}-(\sqrt{g^{-1}\Sigma})_{\mu}^{\;\nu},
\label{K}
\end{eqnarray}
and rectangular brackets stand for traces.
The tensor $\Sigma_{\mu\nu}$ is written in terms of four
St\"uckelberg fields as
\begin{eqnarray}
\Sigma_{\mu\nu}=\partial_\mu\phi^a\partial_\nu\phi^b\eta_{ab},
\quad
\eta_{ab}={\rm diag}(-1, 1, 1, 1).
\end{eqnarray}
$S_{\rm m}$ denotes the action of matter
which is minimally coupled to gravity.

The equations of motion derived from the action are of the form
\begin{eqnarray}
\mpl^2\left(G_{\mu\nu}+m^2 X_{\mu\nu}\right)=T_{\mu\nu},\label{eom}
\end{eqnarray}
where $X_{\mu\nu}$ represents the contribution from ${\cal U}$ due to
the graviton mass and $T_{\mu\nu}$ is the matter energy-momentum tensor.
Here it should be noted that the effective energy momentum
tensor $X_{\mu\nu}$ from ${\cal U}$
is determined only by algebraic manipulation of the
inverse metric matrix $g^{-1}$ in ${\cal K}_{\mu}^{\;\nu}$ defined by
Eq.~(\ref{K}).

In this {\em Letter}, we work in a one-parameter family
of the above theory, where $\alpha_3$ and $\alpha_4$
are given by
\begin{eqnarray}
\alpha_3=\frac{1}{3}(\alpha -1),
\quad
\alpha_4=\frac{1}{12}\left(\alpha^2-\alpha+1\right).
\label{alpha}
\end{eqnarray}
This parameter choice (made also in Ref.~\cite{BHs})
enables us to
find new solutions in massive gravity.
In addition, this choice is
useful for avoiding potential ghost instabilities suggested in
Ref.~\cite{KNT3}, as would-be dangerous fluctuation modes
become nondynamical from the beginning.

Our metric ansatz is taken to be the general PG form~\cite{gPG}:
\begin{eqnarray}
&&\D s^2=-V^2(t,r)\D t^2
+U^2(t, r)\left(\D r+\epsilon\sqrt{f(t,r)}\D t\right)^2
\nonumber\\&&\qquad\qquad
+W^2(t,r)r^2\D\Omega^2,\label{metric-ansatz}
\end{eqnarray}
where $\epsilon=\pm 1$.
We write the St\"{u}ckelberg fields
in the unitary gauge as
\begin{eqnarray}
\phi^0=t,\quad \phi^i=r\hat n^i,\label{stu}
\end{eqnarray}
where $\hat n$ is the unit radial vector,
$\hat n=\left(\cos\varphi\sin\theta, \sin\varphi\sin\theta, \cos\theta\right)$.

In the absence of matter,
the de Sitter solution has been constructed in the coordinate system
of~(\ref{metric-ansatz})~\cite{BHs}:
\begin{eqnarray}
\D s^2=-\kappa^2\D t^2+\tilde\alpha^2\left(\D r\pm\tilde H r\D t\right)^2+\tilde\alpha^2
r^2\D\Omega^2,\label{dS-special}
\end{eqnarray}
where $\tilde H=\kappa m/\sqrt{3\alpha}$, $\tilde \alpha:=\alpha/(\alpha+1)$,
and $\kappa$ is an integration constant.
This solution is different from the de Sitter solution
found by Koyama, Niz, and Tasinato~\cite{KNT1, KNT2}.
In fact, the metric~(\ref{dS-special}) solves the equations of motion
only in the special case where the parameters are given by
Eq.~(\ref{alpha}).

We are going to generalize the work of Ref.~\cite{BHs} to
accommodate a wider class of dynamical solutions
including cosmological ones.
In light of the de Sitter solution~(\ref{dS-special}),
we concentrate on the case satisfying
\begin{eqnarray}
W(t, r)=\tilde\alpha:=\frac{\alpha }{\alpha+1}. \label{W-ansatz}
\end{eqnarray}
The key observation here is that
{\em for any metric of the form~(\ref{metric-ansatz})
with~(\ref{W-ansatz}) and for the St\"{u}ckelberg~(\ref{stu}),
the tensor $X_{\mu\nu}$ reduces to
the effective cosmological constant term:}
\begin{eqnarray}
X_{\mu\nu} =\frac{1}{\alpha} g_{\mu\nu}.\label{Xmn}
\end{eqnarray}
This happens only for the special parameter choice~(\ref{alpha}).
Consequently, {\em any PG-type solution
in general relativity (with a cosmological constant) is also
a solution to massive gravity.}
Equation~(\ref{Xmn}) implies that
the identity
$\mpl^2 m^2\nabla_\mu X^{\mu\nu}=\nabla_{\mu}T^{\mu\nu}-\mpl^2\nabla_{\mu}G^{\mu\nu}=0$
is automatically satisfied.
Our finding thus extends the observations made in Refs.~\cite{Nieu, BHs}
to the general PG metric.

Let us demonstrate how cosmological solutions are obtained
using the above fact.
The FLRW metric can be rewritten in a general PG form as~\cite{Kanai, ABCL}
\begin{eqnarray}
\D s^2&=&-\kappa^2\D t^2+\frac{\tilde \alpha^2}{1-K\tilde\alpha^2r^2/a^2(t)}
\left(\D r-\frac{\dot a}{a}r\D t\right)^2
\nonumber\\&&\qquad
+\tilde\alpha^2 r^2\D\Omega^2,
\label{P-FLRW}
\end{eqnarray}
where $K=0, \pm 1$, a dot denotes differentiation with respect to
$t$, and we have taken $\epsilon =-1$ for expanding flow. In general
relativity, one is free to rescale the time coordinate to remove the
constant $\kappa$. However, this is not the case in massive gravity
because such a rescaling will change the tensor
$\Sigma_{\mu\nu}$, and then $\Sigma_{\mu\nu}$ will depend on $\kappa$.
For this reason, we do not set $\kappa=1$ but rather leave $\kappa$ in
the metric as an integration constant characterizing the solution.

We include a perfect fluid whose energy momentum tensor is given by
\begin{eqnarray}
T_{\mu\nu}=\left(\rho+p\right)u_\mu u_\nu+pg_{\mu\nu},
\label{EMT}
\end{eqnarray}
where $u_\mu=(-V,0,0,0)$.
The energy density and pressure may depend on both $t$ and $r$ in general,
but in the present cosmological setting
they are supposed to depend only on $t$: $\rho=\rho(t), p=p(t)$.

The equations of motion~(\ref{eom}) now read
\begin{eqnarray}
\frac{3}{\kappa^2}
\tilde H^2&=&\frac{\rho}{\mpl^2}+\frac{m^2}{\alpha}-\frac{3K}{a^2},
\label{FRW1}
\\
-\frac{1}{\kappa^2}
\left(3\tilde H^2+2\dot{\tilde H}\right)&=&\frac{p}{\mpl^2}-\frac{m^2}{\alpha}+\frac{K}{a^2},
\label{FRW2}
\end{eqnarray}
where $\tilde H(t):=\D\ln a/\D t$.
From the energy conservation equation, $u_\mu \nabla_\nu T^{\mu\nu}=0$,
we obtain
\begin{eqnarray}
\dot\rho+3\tilde H(\rho + p)=0.
\end{eqnarray}
Rescaling the time coordinate as
\begin{eqnarray}
t\to \tau=\kappa t,\label{tr-t}
\end{eqnarray}
and using $H:=\D\ln a/\D \tau$ instead of $\tilde H$,
one can apparently remove the constant $\kappa$ from the above equations.
Thus, the standard cosmological equations with
the effective cosmological constant
\begin{eqnarray}
\Lambda_{\rm eff}=\frac{m^2}{\alpha}
\end{eqnarray}
are reproduced.
It should be emphasized that in the present case
{\em spatially flat, open and closed models are possible}.
This is in sharp contrast to the findings in Refs.~\cite{Cos, Open1}.
It is straightforward to include a ``bare'' cosmological constant $\Lambda$
by shifting $\Lambda_{\rm eff}\to\Lambda+m^2/\alpha$.

The metric~(\ref{P-FLRW}) may be expressed in a more familiar
``cosmological'' form.  This is done by using the new radial coordinate
defined as
\begin{eqnarray}
r\to\varrho=\frac{\tilde\alpha r}{a(\tau)}.\label{tr-r}
\end{eqnarray}
In terms of $\varrho$, the metric is indeed of
the FLRW form:
\begin{eqnarray}
\D s^2=-\D\tau^2+a^2\left(\frac{\D\varrho^2}{1-K\varrho^2}+\varrho^2\D\Omega^2\right).
\label{usual-FLRW}
\end{eqnarray}
However, this coordinate transformation brings
the St\"{u}ckelberg scalar fields to
\begin{eqnarray}
\phi^0=\frac{\tau}{\kappa},\quad\phi^i=\frac{a(\tau)\varrho}{\tilde\alpha}\hat n^i,
\end{eqnarray}
so that
\begin{eqnarray}
\Sigma_{\mu\nu}\D x^\mu\D x^\nu
&=&-\left(\frac{1}{\kappa^2}-\frac{a^2H^2\varrho^2}{\tilde\alpha^2}\right)
\D\tau^2+2\frac{a^2 H\varrho}{\tilde{\alpha}^2}\D\tau \D\varrho
\nonumber\\&&\qquad 
+\frac{a^2}{\tilde\alpha^2}\left(\D\varrho^2+\varrho^2\D\Omega^2\right).
\end{eqnarray}
Thus, we see that, though the geometry described by the metric~(\ref{usual-FLRW})
is spatially homogeneous and isotropic, the tensor
$\Sigma_{\mu\nu}$ does not respect the same symmetry.
After the coordinate transformation~(\ref{tr-t}) and~(\ref{tr-r}),
$\Sigma_{\mu\nu}$ carries the information about
the constants $\kappa$ and $\tilde\alpha$.

Introducing the St\"{u}ckelberg fields gives rise to
a new invariant $I^{ab}=g^{\mu\nu}\partial_\mu\phi^a\partial_\nu\phi^b$.
It is easy to check that the solution we have obtained shows no singularity
in $I^{ab}$. This owes to the fact that the metric of the PG form
has no coordinate singularity on its horizon.

It would be an important next step to study perturbations around our
cosmological background.  The analysis of cosmological perturbations
will be nontrivial because the reference metric does not respect
the same symmetry as the FLRW one.  However, fluctuations around the de
Sitter background have been investigated in the decoupling limit in
Ref.~\cite{BHs}, and it has been found that the kinetic terms for the
helicity-0 and helicity-$\pm 1$ modes vanish identically.  This is the
consequence of the special parameter choice~(\ref{alpha}). In
particular, as mentioned above, this implies that one can avoid
potential ghost instabilities suggested in Ref.~\cite{KNT3}.

Finally, we mention the description of gravitational collapse in massive
gravity in terms of the PG-type solutions.
Our cosmological metric in the PG coordinates
can also be utilized to describe collapsing matter in
spherically symmetric spacetime
(with $\epsilon=+1$ to adapt the contracting setup).
Indeed, the metric of the PG form has been used to analyze the spherical
contraction model of a star with uniformly distributed
dust~\cite{ABCL}, and with a perfect fluid~\cite{Kanai}, while in the present
case the Schwarzschild-de Sitter solution found in Ref.~\cite{BHs}
is to be used to describe the exterior solution.
In addition, we explicitly show that
the LTB metric representing general inhomogeneous collapsing dust
is also included in the class of our PG-type solutions.


In general relativity, the metric representing spherically symmetric 
spacetime can be expressed in a general PG form without loss of
generality as
\begin{align}
\D s^2=-N^2
\D t^2+\frac{\tilde{\alpha}^2}{1+2E}(N_r \D t+  \D r)^2 +\tilde{\alpha}^2 r^2 \D \Omega^2,
\label{gLTB}
\end{align}
where $N(t,r)>0$ is the lapse function, $N_r(t,r)$ is the radial
component of the shift vector, and $E(t,r)> -1$.
The energy momentum tensor
$T_{\mu\nu}$ is given by Eq.~(\ref{EMT}) while $\rho$ and $p$ are no longer
homogeneous, and a ``bare'' cosmological constant
$\Lambda$ may also be included.
The metric (\ref{gLTB}) can be used to analyze
spherical collapse of a perfect fluid~\cite{gLTB}.

From the equation of motion, we see
\begin{eqnarray}
E=\frac12\left(\frac{\tilde{\alpha}N_r}{N}\right)^2 -\frac{M}{\tilde{\alpha}r},\label{Energy}
\end{eqnarray}
where we defined an enclosed mass
\begin{eqnarray}
M(r,t):=4\pi\int^r \left(\rho+\mpl^2\Lambda\right)
{r'}^2 \D r'.
\end{eqnarray}
Here, $E$ represents energy,
which is conserved for dust, as shown later.

The analytic solution for general inhomogeneous dust collapse is most
commonly expressed in the LTB coordinates. We change the
coordinates from $(t,r,\theta,\phi)$ to $(T,R,\theta,\phi)$ such that
$t=t(T) =T$ and $r=r(T,R)=\bar{r}(T,R)/\tilde{\alpha}$ with
\begin{align}
\left(\frac{\partial \bar{r}}{\partial T}\right)
=-\tilde{\alpha}N_r=-N\sqrt{\frac{2 M}{\bar{r}}+2E},
\end{align}
where we have taken only the positive root of Eq.~(\ref{Energy})
for a collapsing fluid.
Now, $N, M$, and $E$ are all functions of $T$ and $R$, and
the metric is of the LTB form:
\begin{align}
\D s^2 = -N^2 \D T^2+\frac1{1+2E}\left(\frac{\partial \bar{r}}{\partial R}\right)^2 \D R^2
+\bar{r}^2 \D \Omega^2.
\label{LTB}
\end{align}
In the case of dust in which one can eliminate the pressure gradient from
the relevant equations,
$N$ is independent of $R$ and hence is
allowed to be synchronous, $N=1$.
However, for the application to massive gravity,
it would be better to leave $N$ in the metric
as an additional arbitrary function
of $T$.
Moreover, it turns out that
$E$ is independent of $T$, and then the remaining
evolution equations have three types of solutions
depending on the sign of $E(R)$.
This is the well known LTB solution that can be expressed
in the PG form (\ref{gLTB}) as well.

On the other hand, in massive gravity with the parameter choice~(\ref{alpha}),
one immediately sees that the LTB solution in the PG form~(\ref{gLTB})
with a cosmological constant $\Lambda+m^2/\alpha$
is the solution to~(\ref{eom}) in the unitary gauge.
This is because,
as
mentioned above, given metric functions of the generalized PG form, the
effective energy momentum tensor $X_{\mu\nu}$ of ${\cal U}$
reduces to the
cosmological term, $(1/\alpha) g_{\mu\nu}$, irrespective of their
coordinate dependence.
The coordinate transformation from PG to LTB
with the rescaling $T\to T/N$
brings the
St\"{u}ckelberg scalar fields to
\begin{eqnarray}
\phi^0=\int \frac{\D T}{N},
\quad
\phi^i=\frac{\bar{r}}{\tilde{\alpha}}\hat n^i,
\end{eqnarray}
so that
\begin{eqnarray}
\Sigma_{\mu\nu}\D x^\mu \D x^\nu=-\frac{\D T^2}{N^2}+
\left(\frac{\partial\bar{r}}{\partial T}\frac{\D T}{\tilde{\alpha}}
+\frac{\partial\bar{r}}{\partial R}\frac{\D R}{\tilde{\alpha}} \right)^2
+\left(\frac{\bar{r}}{\tilde{\alpha}} \right)^2\D\Omega^2.
\end{eqnarray}
Note that also in this case the invariant $I^{ab}$ does not
diverge on the horizon by virtue of the PG coordinate system.

In summary, we have found new cosmological solutions in massive gravity with
flat, open, and closed spatial geometries.
Our solutions can also
describe inhomogeneous gravitational collapse of dust represented by
the LTB metric. The key was that
general PG-type metric gives rise to an effective energy momentum tensor
of a cosmological
constant $m^2/\alpha$ for the special choice of the parameters.
This is essential to hunt for analytical
solutions in massive gravity from the seed solutions in general
relativity.
Thus, our solutions can be used not only for
homogeneous and isotropic cosmology with arbitrary spatial curvature,
but also for the spherical collapse model of the formation of cosmic structure
such as stars and galaxies.


In this {\em Letter}, the special
choice of the parameters of the theory enabled us to have
the desirable structure $X_{\mu\nu}\propto g_{\mu\nu}$.
Consequently, the conservation law $\nabla_\mu X^{\mu\nu}=0$
is satisfied automatically.
The presence of the conservation law suggests the presence of
some accidental symmetry in the PG-type metric.
It would be thus interesting to understand more deeply the nature of
the metrics giving the effective cosmological term.
As a concrete example, we will report the behavior of
perturbations on our cosmological background in a separate publication~\cite{toappear}.\\


\paragraph*{Acknowledgments}

We would like to thank Takeshi Chiba, Jiro Soda, and Daisuke Yamauchi for
useful discussions. This work is supported in part by JSPS Grant-in-Aid
for Young Scientists (B) No.~24740161 (T.K.), No.~21740187 (M.Y.), and
Scientific Research on Innovative Areas No.~24111706 (M.Y.).\\


{\bf Note added~~} While this paper was being completed,
Ref.~\cite{Gratia} appeared,
in which a similar situation is discussed
where an effective cosmological term
$X_{\mu\nu}\propto  g_{\mu\nu}$ arises.
The authors of Ref.~\cite{Gratia} derived
a constraint
that is to be imposed on the metric and the St\"{u}ckelberg functions,
while we have given explicit solutions
for the metric and the St\"{u}ckelberg fields
in a simple manner.



\end{document}